\documentclass[aps,pra,superscriptaddress,twocolumn]{revtex4-1}
\usepackage[utf8]{inputenc}
\usepackage{amsmath}
\usepackage{amsfonts}
\usepackage{mathtools}
\usepackage{physics}
\usepackage{dsfont}
\usepackage{verbatim}
\usepackage{color}
\usepackage[dvipsnames]{xcolor}
\usepackage{soul}
\usepackage[utf8]{inputenc}
\usepackage[ruled,longend]{algorithm2e}
\usepackage{textgreek}
\usepackage{amssymb}
\usepackage[normalem]{ulem}

\usepackage{tcolorbox}

\newcommand{\nico}[1]{\textcolor{blue}{\textit{#1}$^{\rm Nico}$}}

\newcommand{\cmnt}[1]{}

\begin{document}

\title{A hybrid classical-quantum approach to speed-up Q-learning}

\author{A. Sannia}
\affiliation{Dipartimento di Fisica, Universit\`a della Calabria, 87036
	Arcavacata di Rende (CS), Italy}
\author{A. Giordano}
\affiliation{ICAR-CNR, 87036 Rende, Italy}
\author{N. Lo Gullo}
\affiliation{Quantum algorithms and software, VTT Technical Research Centre of Finland Ltd}
\affiliation{CSC-IT Center for Science, P.O. Box 405, FIN-02101 Espoo, Finland}
\author{C. Mastroianni}
\affiliation{ICAR-CNR, 87036 Rende, Italy}
\author{F. Plastina}
\affiliation{Dipartimento di Fisica, Universit\`a della Calabria, 87036 Arcavacata di Rende (CS), Italy}
\affiliation{INFN, gruppo collegato di Cosenza}

\begin{abstract}
We introduce a classical-quantum hybrid approach to computation, allowing for a quadratic performance improvement in the decision process of a learning agent. In particular, a quantum routine is described, which encodes on a quantum register the probability distributions that drive action choices in a reinforcement learning set-up. This routine can be employed by itself in several other contexts where decisions are driven by probabilities. After introducing the algorithm and formally evaluating its performance, in terms of computational complexity and maximum approximation error, we discuss in detail how to exploit it in the Q-learning context.
\end{abstract}

\maketitle

\section{Introduction}

%Nuovi riferimenti su QML/QRL: \cite{biamonte2017}, \cite{dunjko2016}, \cite{saggio2021}, \cite{sriarunothai2018}, \cite{jerbi2021}, \cite{porotti2019}, \cite{paparo2014}

Quantum algorithms can produce statistical patterns not so easy to obtain on a classical computer; in turns,  they may, perhaps, help recognize patterns that are difficult to identify classically. To pursue this basic idea, a huge research effort is being put forward to speed up machine learning routines by exploiting unique quantum properties, such as coherence and entanglement, ultimately due to the superposition principle, \cite{biamonte2017,dunjko2016}.

Within the realm of machine learning, the Reinforcement Learning (RL) paradigm has gained huge attention in the last two decades \cite{RL-survey,sutton2018}, as, in a wide range of application scenarios, it allows modeling an agent that is able to learn and improve its behavior through rewards and penalties received from a not fully known environment. The agent, typically, chooses the action to perform by sampling a probability distribution that mirrors the expected returns associated to each of the actions possibly taken in a given situation (state). The estimated outputs and the corresponding probability distribution for the actions need to be updated at every step. 

It turns out that the efficiency of RL may be improved by the use of hybrid classical-quantum   \cite{Dong,QRL-Nature,QRLPhotonics,jerbi2021variational,quantum2020019,shenoy2020demonstration}, where, e.g., quantum routines can help to prepare/update the probability distributions that drive the agent operations \cite{saggio2021,PhysRevX.4.031002,sriarunothai2018}. 

In this paper, we provide a novel algorithm for updating a probability distribution on a quantum register, which does not require the knowledge of the probabilities for all of the admissible actions, as assumed in other works \cite{GroverPDF,paperJPMorgan}. Instead, the actions are clustered in a predetermined number of subsets (classes), each associated to a range with a minimum and a maximum value of the expected reward. The cardinality of each class is evaluated in due course, through a procedure built upon well-known quantum routines, i.e., quantum oracle and quantum counting \cite{Nielsen,counting}. Once this information is obtained, a classical procedure is run to assign a probability to each subset, in accordance to any desired distribution, while the elements within the same class are taken to be equally likely. This allows one to tune probabilities, in order, e. g., to assign a large one to the actions included in the range with maximum value of the expected reward. The probability distribution can also be changed dynamically, in order to enforce exploration at a first stage (to allow choosing also actions that have a low value) and exploitation at a second stage (to restrict the search to actions having a high value). The quantum computing algorithm presented here allows re-evaluating the values after examining all the actions that are admissible in a given state in a single parallel step, which is only  possible due to quantum superposition. 

Besides the RL scenario, for which our approach is explicitly tailored, the main advantageous features of this algorithm
 %(\textbf{vedi alla fine della sezione})
 could be also exploited  in several other contexts where one needs to sample from a distribution, ranging from swarm intelligence algorithms (such as Particle Swarm Optimization and Ant Colony Optimization \cite{BookSwarm}), to Cloud architectures (where the objective is to find an efficient assignment of virtual machines to physical servers, a problem that is known to be NP-hard \cite{EcoCloud}).
After presenting the algorithm in details in Sec. \ref{secuno} (while postponing to the appendix a formal evaluation of its performance in terms computational complexity and maximum approximation error), we will focus on its use in the RL setting in Sec. \ref{secdue}, to show that it is, in fact, tailored for the needs of RL with a finite number of actions/states of the environment. We, finally draw some concluding remarks in Sec. \ref{sec:conclu}.

\section{Preparing a Quantum Probability Distribution} \label{secuno}
We now present the main building blocks of our approach to encode a user-defined probability distribution into a Quantum Register (QR). This procedure may turn useful in every application where it is required to extract the value of a random variable.

Let us assume that we have a random variable whose discrete domain includes $J$ different values, $\{x_j:j=1, \ldots, J\}$, which we map into the basis states of a $J$-dimensional Hilbert-space. Our goal is to prepare a quantum state for which the measurement probabilities in this basis reproduce the random variable probability distribution: $\{p_{x_1},p_{x_2},\cdots,p_{x_{J}}\}$.

The algorithm starts by initializing the QR as:
\begin{align}
\label{eq:initial}
\ket{\psi}=\frac{1}{\sqrt{J}} \sum_{k=1}^{J}\ket{x_k}\ket{1}_a  \equiv \ket{\phi} \ket{1}_a\ 
\end{align}
where $\ket{\phi}$ is the equal superposition of the basis states $\{\ket{x_k}\}$, while an ancillary qubit is set to the state $\ket{1}_a$.
In our approach, the final state is prepared by encoding the probabilities sequentially, which will require $J-1$ steps. 

At the $i$-th step of the algorithm ($1\leq i < J$), Grover's iterations \cite{grover} are used to set the amplitude of the $\ket{x_i}$ basis state to $a_i=\sqrt{p_{x_i}}$.
In particular we apply a conditional Grover's operator :  $\mathbb{I}\otimes\vu*{\Pi}_{a}^{(0)} + \vu*{G}_i\otimes \vu*{\Pi}_{a}^{(1)}$, where  $\vu*{G}_i=\vu*{R}\vu*{O}_i$ and $\vu*{\Pi}_{a}^{(y)}$ is the projector onto the state $\ket{y}_a$ of the ancilla ($y=0,1$). It forces the Grover unitary, $\vu*{G}_i$, to act only on the component of the QR state tied to the ('unticked') state $\ket{1}_a$ of the ancillary qubit. The operator $\vu*{R}=2\ket{\phi}\bra{\phi}-\mathbb{I}$ is the reflection with respect to the uniform superposition state, whereas the operator $\vu*{O}_i=\mathbb{I}-2\ket{x_i}\bra{x_i}$ is built so as to flip the sign of the state $\ket{x_i}$ and leave all the other states unaltered. The Grover's operator is applied until the amplitude of $\ket{x_i}$ approximates $a_i$ to the desired precision (see Appendix \ref{sec:iteration}). The state of the ancilla is not modified during the execution of the Grover's algorithm.

The $i$-th step ends by ensuring that the amplitude of $\ket{x_i}$ is not modified anymore during the next steps. To this end, we 'tick' this component by tying it to the state $\ket{0}_a$. This is obtained by applying the operator $\vu*{F}_i=\ket{x_i}\bra{x_i}\otimes\vu*{X}+ \big(\mathbb{I}-\ket{x_i}\bra{x_i}\big)\otimes \mathds{1} $, whose net effect is: 
\begin{equation*}
\vu*{F}_i\ket{x_k}\ket{y}_a=\begin{cases} \ket{x_k}\otimes \vu*{X}\ket{y}_a , & \mbox{if $k=i$ } \\ \ket{x_k}\ket{y}_a, & \mbox{otherwise}
\end{cases}
\end{equation*}
Where  $\vu*{X}$ is the NOT-gate, with $1\leq i \leq J$, $y \in \{0,1\}$.\\

After the step $i$, the state of the system is: 
\begin{eqnarray*}
\ket{\psi}&=& \sum\limits_{j=1}^{i} a_j \ket{x_j}\otimes \ket{0}_a +  b_{i} \ket{\beta_{i}}\ket{1}_a.
\end{eqnarray*}
The state $\ket{\beta_{i}}$ has in general non-zero overlap with all of the basis states, including those states, $\ket{x_1}$,$\cdots$,$\ket{x_{i-1}}$, whose amplitudes have been updated previously. This is due to the action of the reflection operator $\vu*{R}$, which outputs a superposition of all the $J$ basis states.
However, this does not preclude us from extracting the value of the probability of the random variable correctly, thanks to the ancillary qubit. Indeed, at the end of the last step, the ancilla is first measured in the logical basis. If the outcome of the measurement is $0$, then we can proceed measuring the rest of the QR to get one of the first $J-1$ values of the random variable with the assigned probability distribution. Otherwise, the output of the algorithm is set to $x_J$, since the probability of getting $1$ from the measurement of the ancilla is $p_{x_{J}}$ due to the normalization condition ($p_{x_{J}}=a_{J}^2 = 1- \sum_{k=1}^{J-1} a_k^2$).
\cmnt{
To illustrate how the algorithm works, let us look at the first step. The state of the system evolves as:
\begin{align*}
&\ket{\psi}=\ket{\phi}\ket{1}_a, &\text{(initial state)}\\
&\ket{\psi}=a_1 \ket{x_1}\ket{1}_a+b_1 \ket{\beta_1}\ket{1}_a, &\text{(Grover's iterations applied)}\\
&\ket{\psi}=a_1 \ket{x_1}\ket{0}_a+b_1 \ket{\beta_1}\ket{1}_a, &\text{(state $\ket{x_1}$ marked)}
\end{align*}
where $\ket{\beta_1}=\frac{1}{\sqrt{J-1}}\sum_{i=2}^J \ket{x_i}$, and the  values of $a_1$ and $b_1$ depend on the number of iterations of $\vu*{G}_1$. Notice that, as mentioned before,  the state $\ket{x_1}$ is correlated with the $\ket{0}_a$ state of the ancilla to ensure that further applications of the conditional Grover's operator will not modify the amplitude $a_1$.}

The procedure can be generalized to encode a random distribution for which $N$ elements, with $N>J$, are divided into $J$ sub-intervals. In this case, a probability is not assigned separately to every single element; but, rather, collectively to each of the $J$ sub-intervals, while the elements belonging to the same sub-interval are assigned equal probabilities. This is useful to approximate a random distribution where the number of elements, $N$, is very large. In this case, the QR operates on an $N$-dimensional Hilbert space, and Grover's operators are used to amplify more than one state at each of the $J-1$ steps. For example, in the simplest case, we can think of having a set of $J$ Grover operators, each of which acts on $N/J$ basis states. In the general case, though, each Grover's operator could amplify a different, and a priori unknown, number of basis states. This case will be discussed in the following where the algorithm is exploited in the context of RL. 

%Let us consider a random variable which can take on $N$ possible values of a random variable, with $N$ being a large number. We first need to map the values of the random variable onto a QR and then find a strategy to encode the values of the probability distribution into it.

%Mapping the values of the random variables would require a set of basis states (therefore orthogonal and normalized by definition) $\{\ket{k}\}$ with cardinality of the order of $N$. This can be obtained by means of a QR made of $\approx \log_2N$ qubits resulting, in general, in a large Hilbert space. Due to the limitation in both number of qubits and maximum depth of quantum circuits on NISQ processors we propose a way to reduce the number of required states.

%{\it Improving Reinforcement Learning} - 
\section{Improving Reinforcement Learning} \label{secdue}
\begin{figure}
    \centering
    \includegraphics[width=\linewidth]{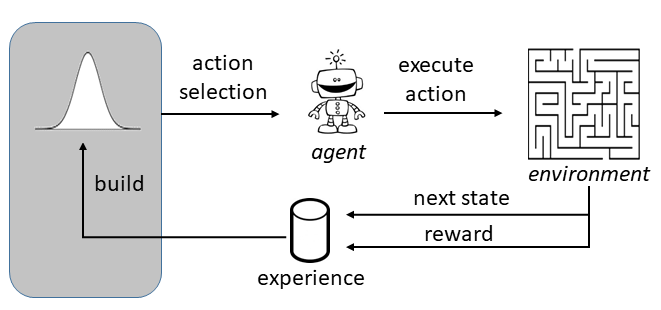}
    \caption{Representation of a Q-learning cycle, where the quantum advantage is brought (on the left part) by the algorithm that encodes and updates action probabilities on a QR.}
    \label{fig:workflow}
\end{figure}

We will now show how the algorithm introduced above can be exploited in the context of RL, and, specifically, in the Q-learning cycle. Figure \ref{fig:workflow} provides a sketch emphasising the part of the cycle that is involved in our algorithm. Our objective, in this context, is to update the action probabilities, as will be clarified in the rest of this section.  
An RL algorithm can be described in terms of an abstract \textit{agent} interacting with an \textit{environment}. The agent can be in one of the \textit{states} that belong to a given set $S$, and is allowed to perform \textit{actions} picked from a set $A_s$, which, in general, depends on $s$. For each state $s \in S$, the agent chooses one of the allowed actions, according to a given policy. After the action is taken, the agent receives a \textit{reward} $r$ from the environment and its state changes to $s' \in S$. The reward is used by the agent to understand whether the action has been useful to approach the goal, and then to learn how to adapt and improve its behavior. Shortly, the higher the value of the reward, the better the choice of the action $a$ for that particular state $s$. In principle, this behavior, or \textit{policy}, should consist in a rule that determines the best action for any possible state.
An RL algorithm aims at finding the optimal policy, which maximizes the overall reward, i.e., the sum of the rewards obtained after each action. However, if the rewards are not fully known in advance, the agent needs to act on the basis of an estimate of their values.

Among the various approaches designed to this end, the so called Q-Learning algorithm \cite{sutton2018} adopts the Temporal Difference (TD) method to update  
%two quantities: the $V(s)$ value, i.e., an estimation of the convenience to move to the state $s$, and
the $Q(s,a)$ value, i.e., an estimation of how profitable is the choice of the action $a$ when the agent is in the state $s$.

When choosing an action, at a given step of the algorithm, two key factors need to be taken into account: \textit{explore} all the possible actions, and \textit{exploit} the actions with the greatest values of $Q(s,a)$. As it is common, we resort to a compromise between exploration and exploitation by choosing the new action through a random distribution, defined so that the probability of choosing the action $a$ in the state $s$ mirrors $Q(s,a)$. For example, one could adopt a Boltzmann-like distribution $
P(a|s)=e^{Q(s,a)/T}/Z$, where $Z$ is a normalising factor, while the $T$ parameter can vary during the learning process (with a large $T$ in the beginning, in order to favor exploration, and lower $T$ once some experience about the environment has been acquired, in order to exploit this knowledge and give more chances to actions with a higher reward).

A severe bottleneck in the performance of a TD training algorithm arises when the number of actions and/or states is large. For example, in the chess game, the number of states is $\sim 10^{120}$ and it is, in fact, impossible to deal with the consequent huge number of $Q(s,a)$ values. A workaround is to use a function $Q^{*}_{\vec{\theta}}(s,a)$ that \textit{approximates} the values of $Q(s,a)$ obtained by the TD rule and whose properties depend upon a (small) set of free parameters $\vec{\theta}$ that are updated during the training. This approach showed its effectiveness in different \textit{classical} approaches as, for example, in Deep Q-Learning \cite{dqn2018,van2016deep,dulac2015deep,8402196,andriotis2019managing}, where the $Q^{*}_{\vec{\theta}}(s,a)$ function is implemented by means of a neural network whose parameters are updated in accordance with the experience of the agent.

%\textst{In the following, we will assume to work with a conditional probability $P_{\vec{\theta}}(a|\bar s)=N_a \cdot f(Q^{*}_{\vec{\theta}}(s,a))$, namely the probability that once the state $s$ is observed, the agent will perform the action $a$, where $N_a$ is a renormalization factor. Ideally, this probability will tend to been chosen by rharpen around the most profitable $a$ at the end of the training.}

In the quantum scenario, this approach turns out to be even more effective. Indeed, it is possible to build a parameter-dependent quantum circuit that implements $Q^{*}_{\vec{\theta}}(s,a)$; an approach that has been adopted by recent studies on near-term quantum devices \cite{jerbi2021,skolik2021quantum,chen2020variational,He_2021,lockwood2020reinforcement}. This circuit allows us to evaluate the function $Q^{*}_{\vec{\theta}}(s,a)$ in a complete \textit{quantum parallel} fashion; i.e., in one shot for all the admissible actions in a given state.
With this approach, it is possible to obtain a quantum advantage in the process of building the probability distribution for the actions, using the algorithm presented in Sec. ~\ref{secuno}.
To achieve a significant quantum speed-up, and reduce the number of required quantum resources, thus making our algorithm suitable for near term NISQ processing units, we do not assign a probability to every action; but, rather, we aggregate actions into classes (i.e., subsets) according to their probabilities, as explained in the following. Let us consider the minimum ($m$) and maximum ($M$) of the $Q^*_{\vec{\theta}}(s,a)$ values,
 %\footnote{The minimum and maximum values can be efficiently computed with well-known quantum algorithm (\textbf{citazione})}
 and let us divide the interval $[m,M]$ into $J$ (non-overlapping, but not necessarily equal) sub-intervals $I_j$ with $1 \le j \le J$.
For a given state $s$, we include the action $a\in A_{s}$ in the class $C_j$ if $Q^*_{\vec{\theta}}(s,a) \in I_j$ (so that, $A_s = \bigcup_j C_j$). 

The probability of each sub-interval, then, will be determined by the sum of the $Q^*$-values of the corresponding actions, $\sum_{a \in C_j} Q^*_{\vec{\theta}}(s,a)$. All of the actions in $C_j$, will be then considered equally probable.

In this way, our algorithm requires only $J-1$ steps,
each of which is devoted to amplify the actions belonging to one of the $J-1$ classes (while the $J$-th probability is obtained by normalization). Furthermore, we can also take advantage of the aggregation while encoding the probability distributions onto the QR: in this case, indeed, we can use predetermined Grover's oracles, each devoted to amplify the logical states corresponding to the actions belonging to a given $C_j$. 

\cmnt{
Ideally this circuit takes as an input an homogeneous superposition of the states encoding all allowed actions for a given state $s$ and changes the amplitude of each (action-encoding) state according to the rewards (penalty) received. An action is then picked by projecting onto the action space.
This is to be compared with the classical case where the computation of $Q^{*}_{\vec{\theta}}(s,a)$ for all the actions $a$ occurs sequentially. Moreover in the classical case, to find a good balance between exploitation (picking the most rewarding action) and exploration (try new actions even if they correspond to a low reward) has to be implemented. The quantum circuit-implementation offer this for free due to the probabilistic nature of the measurement.}

In order to insert the distribution-update algorithm into the Q-learning procedure, we need two QRs: $\mathcal{A}$ and $\mathcal{I}$, encoding the actions and the sub-intervals, respectively. These registers need $\lceil log_2(\max_{s}|A_s|)\rceil$ and $\lceil log_2(J)\rceil$ qubits, respectively.
%, where $A_s$ is the set of
%{\it all}
%admissible actions when the agent is in the %state $s$.
Let us consider the class $C_j=\{a \in A_s : Q^{*}(s,a) \in I_j\}$.
%
%We also define $\vu*{O}_{b_i}=\mathbb{I}-2\ket{a'_i}\bra{a'_i}$, where $\ket{a'_i}=\frac{1}{\sqrt{N_i}}\sum_{j}\ket{a_j}$with $a_j \in A_i$.
%
Our goal is to assign to each action $a \in A_j$ a probability $p_j$, based on the sub-interval $j$, using the algorithm presented in Sec.~\ref{secuno}.
The distribution building process starts by preparing the following uniform superposition:
\begin{align*}
\label{eq:initialsuperposition}
\ket{\psi_{s}}=\frac{1}{\sqrt{|A_{s}|}}\sum_{a \in A_{s}}\ket{a}_{\mathcal{A}}\ket{0}_{\mathcal{I}},
\end{align*}
where $\ket{0}_{\mathcal{I}}$ is the initial state of the register $\mathcal{I}$. 
In order to apply our algorithm, we need $J-1$ oracles $\vu*{O}_j$, one for each given $I_j$. 
To obtain these oracles, it is first necessary to define an operator $\vu*{J}$ that records the sub-interval $j_a$ to which the value $Q^{*}_{\vec{\theta}}(s,a)$ belongs. Its action creates correlations between the two registers by changing the initial state of the $\mathcal{I}$- register as follows:
\begin{equation*}
\label{eq:fasce}
\vu*{J} \ket{a}_{\mathcal{A}}\ket{0}_{\mathcal{I}} = \ket{a}_{\mathcal{A}}\ket{j_a}_{\mathcal{I}}. 
\end{equation*}
To complete the construction of the oracles, we need to execute two unitaries: i) the operator $\vu{O'}_{j_a}=\mathbb{I}_\mathcal{A} \otimes(\mathbb{I}_\mathcal{I}-2\ket{j_a}\bra{j_a})$, which flips the phase of the state $\ket{j_a}$ of the $\mathcal{I}$-register; and, ii) the operator $\vu*{J}^{\dagger}$, which disentangles the two registers. 
The effective oracle operator entering the algorithm described in the previous section is then defined as $\vu O_{j}=\vu*J^{\dagger}\vu O'_{j}\vu*J$, its net effect being

\begin{equation*}
\vu*{O}_j \ket{a}_\mathcal{A}\ket{0}_\mathcal{I}=\begin{cases} -\ket{a}_\mathcal{A}\ket{0}_\mathcal{I}, & \mbox{if a $\in$ $C_j$ } \\ \ket{a}_\mathcal{A}\ket{0}_\mathcal{I}, &  \mbox{otherwise}
\end{cases}
\end{equation*}
Eventually, we apply the reflection about average $\vu*R$ on the register $\mathcal{A}$, thus completing an iteration of the Grover operator.

If the cardinality of each $C_j$ is not decided from the beginning, in order to evaluate the right number of Grover iterations to be executed, we need to compute it (see App. \ref{app:opt_num} for details). This number of actions can be obtained, for each $C_j$, by running the quantum counting algorithm associated with $\vu*{O}_j$ (and before its action). It is then possible to apply the algorithm of Sec.~\ref{secuno} in order to build the desired probability distribution.

We stress that our procedure works for any assignment rule of the probabilities to sub-intervals, and the Boltzmann distribution mentioned above is just one example providing a good balance between exploration and exploitation \cite{Boltz,CLOUSE199292}.

After the quantum state of the $\mathcal{A}$-register is obtained, the agent will choose the action measuring its state. Then, according to the outcome of the environment, it will update the $\vec{\theta}$ values classically, thus changing the behaviour of the operator $\vu*{J}$. 

In order to give an overall summary of the procedure we are proposing, the full process, with the various steps we have described, is summarized in the box below, where classical and quantum operations are denoted as (C) and (Q), respectively.

\begin{tcolorbox}[fonttitle=\bfseries, title=Scheme of the hybrid algorithm]
\label{box:QEQ-l}
Initialize $\vec{\theta}$ and start from state $s$ (C)\\
Execute the cycle:
\begin{itemize}
%\item Compute max and min values of $Q^*_{\vec{\theta}}(s,a)$ for each admissible action $a$ (Q)
\item Build the sub-intervals, classes and the quantum circuits for the oracles $\vu*{O}_i=\vu*J^{\dagger}\vu O'_{i}\vu*J$ (C)
\item Use quantum counting on $\vu*{O}_i$ to compute the number of actions belonging to each sub-interval (Q)
\item Compute the number of Grover iterations for each class (C)
\item Build the probability distribution for the admissible actions in $s$ (Q)
\item Measure, obtain an action and execute it (C)
\item Get the new state $s'$ and the reward $r$ (C)
\item Update $\vec{\theta}$ (C) 
\end{itemize}
\end{tcolorbox}

Before concluding, we determine the advantage that can be obtained with our quantum algorithm. Indeed, to build the probability distribution classically, for a given state $s$, the number of calls of the function $Q^{*}_{\vec{\theta}}(s,a)$ increases asymptotically as $\textbf{O}(|A_{s}|)$. Conversely, with our quantum protocol the number of calls of $\vu*{J}$, and therefore of $Q^{*}_{\vec{\theta}}(s,a)$, is asymptotically
$\textbf{O}(\sqrt{|A_{s}|})$.
Our algorithm is devised for the case $|A_{s}|>>1$, which also corresponds to a good precision (see Appendix \ref{app:opt_num} for details). Finally, we point out that in the process of the definition of sub-intervals, it is possible to compute the maximum (M) and the minimum (m) value of $Q^{*}_{\vec{\theta}}(s,a)$ before running our algorithm trough quantum routines that do not increase the complexity of our procedure \cite{MAX, MIN}.

\section{Conclusions}
\label{sec:conclu}
In our work, we presented an algorithm, based on the Grover's one, to encode a probability distribution onto a quantum register with a quadratic speed-up improvement. This algorithm can find several useful applications in the context of hybrid classical-quantum workflows. In this spirit, we have shown how such an algorithm can be exploited for the training of the  Q-learning strategy. We have shown that this gives rise to a quadratic quantum speed up of the RL algorithm, obtained by the inclusion of our quantum subroutine in the stage of action selection of the RL workflow. This effectively enables achieving a trade off between exploration and exploitation, thanks to the intrinsic randomness embodied by the extraction from a QR of the action to be performed and, also, to the possibility of dynamically changing the relationship between the action and their values (and, thus, their relative probabilities). The latter, in particular, would become too burdensome in a fully classical setting. 

Finally, we stress once again that, with our procedure, we can use Grover's oracles, which are given once and for all if i) the minimum and maximum range of action values,  and ii) the number of intervals in which this range is dived are specified.
%\here

\bibliographystyle{IEEEtran}
\bibliography{references}

%\newpage

\appendix
\section{Single step of the probability distribution quantum encoding algorithm}
\label{sec:iteration}

In this appendix we provide some details on a single step of the algorithm presented in the main text, which are important for its actual implementation. Specifically, we address the problems of how to compute the optimal number of iterations of the Grover algorithm to store a single instance of the probability distribution and how to compute quantities which are needed to link the update of the values of the probability distribution on different sub-intervals from one step to another.

\subsection{Optimal Number of iterations of the Grover algorithm}
\label{app:opt_num}
In order to compute the optimal number of iterations in a single step we exploit the results reported in Ref.~\cite{GeneralizedGrover}, where the Grover algorithm has been generalized to the case of an initial non-uniform distribution and define the following quantities :
\begin{align*}
\bar K^{(i)}(t)=\frac{1}{r_{i}} \sum_{j=1}^{r_{i}} k_j^{(i)}(t)\\
\bar L^{(i)}(t)=\frac{1}{N-r_{i}} \sum_{j=r_{i}+1}^{N} l_j^{(i)}(t)
\end{align*}
where t is the number of Grover iteration already performed, N is the dimension of the Hilbert space (in our context it is the total number of actions : $N=|A_s|$) , $\{k_j^{(i)}(t)\}$ are the coefficients of the $r_i$ basis states that will be amplified by the Grover iterations at the step $i$ of the algorithm (in our application $r_i=|C_i|$), and $\{l_j^{(i)}(t)\}$ are the coefficients of all the other basis states, while $\bar K^{(i)}(t)$ and $\bar L^{(i)}(t)$ are their averages and we have labeled them with the step-counting variable $i$.
Let's assume now that only one basis state at a time is amplified by Grover iterations, namely $r_i=1$.
Applying the results in Ref.~\cite{GeneralizedGrover} to our case we obtain:

\begin{eqnarray}
\label{eq:k_L}
\bar K^{(i)}(t) &= k^{(i)}(0)\cos(wt)+\bar L^{(i)}(0)\sqrt{N-1}\sin(wt)\nonumber\\
&\\
\bar L^{(i)}(t)&=\bar L^{(i)}(0)\cos(wt)-k^{(i)}(0)\sqrt{\frac{1}{N-1}}\sin(wt)\nonumber,
\end{eqnarray} 
where $w=2\arcsin(\sqrt{1/N})$.
With the first of these equations we can compute the number of steps $t_f^{(i)}$ needed to set the coefficient $k(t)$ to the desired value with the wanted precision so as to bring the value of the probability distribution to $P(x_i)=|b_i\;k^{(i)}(t_f^{(i)})|^2$.
Notice that we need the values of $k^{(i)}(0)$ and $\bar L^{(i)}(0)$ to perform this calculation, which values can be extracted from the form of the global state at the previous step and specifically from the last iteration of the Grover algorithm.

\subsection{Variation of quantum state within the Grover iterations}

Let us consider the quantum state of the action-register plus the ancilla system at a given iteration $t$ of the Grover algorithm at a given step $i$:
\begin{align*}
\ket{\psi(t)}=\sum_{k=1}^{i-1}a_k\ket{x_k}\ket{0}_a+b_i\ket{\beta_i(t)}\ket{1}_a
\end{align*}
where $ b_i=(1-\sum_{k=1}^{j-1}a_k^2)^{1/2}$ and $t=0$ at the beginning of the Grover algorithm.
Let us write the state $\ket{\beta_i(t)}$ in a form which highlights its decomposition into three sets of basis states: 
\begin{equation}
\ket{\beta_i(t)}=k^{(i)}(t)\ket{x_i} + \sum_{k=1}^{i-1}l_k^{(i)}(t)\ket{x_k}+\sum_{k=i+1}^{N}\alpha^{(i)}(t)\ket{x_k},
\end{equation}
where we have made explicit the dependence of the coefficients of the decomposition on the step $i$, for this will be useful in the following.
The $\ket{x_i}$ is the one we want to use to encode the value of the probability distribution at the current step $i$ of our algorithm, the $\{x_j\}$ with $j\in[1,i-1]$ basis states that are generated by the reflection operation around the mean and whose amplitudes have been updated in the previous $i-1$ steps, and the $\{x_{j}\}$ with $j\in[i+1,N]$ basis state, all having the same amplitude $\alpha^{(i)}(t)$ for all the operations performed up to this point did not change them.

%Recall that the controlled Grover operator $\vu*{G}_i$ acts non-trivially only onto the state $\ket{\beta_i(t)}$, because of the projector onto the acilla's state $\ket{1}_a$

Using this expression it is possible to derive a recursive relation to compute $\alpha^{(i)}(t)$ iteratively as a function of $\bar{L}^{(i)}(t)$ and $k^{(i)}(t)$. As we shall see below this will be useful in order to compute the initial $k^{(i+1)}(0)$,$\bar L^{(i+1)}(0)$ and  $\alpha^{(i+1)}(0)$ for the next step. 
To find this recursive relation let us first apply the Grover operator onto $\ket{\beta_i(t)}\ket{1}_a$ and then project onto any of the states $\{x_{i+1}, x_{i+2}, \cdots, x_{N}\}$. Without loss of generality we choose $x_{i+1}$:

\begin{eqnarray} 
\label{eq:alpha}
&\;\;&\alpha^{(i)}(t+1) =  \bra{x_{i+1}}\vu*{R}\vu*{O}_i\ket{\beta_i(t)} \\
&= & \bra{x_{i+1}}\Bigl(2\ket{\phi}\bra{\phi}-\mathbb{I}\Bigr)\times\nonumber\\
&&\left(-k^{(i)}(t)\ket{x_i}+\sum_{k=1}^{i-1}l_k^{(i)}\ket{x_k} + \sum_{k=i+1}^{N}\alpha^{(i)}(t)\ket{x_k}\right)
\nonumber \\ &=& \bra{x_{i+1}}\Bigl \{2\left(-\frac{1}{\sqrt{N}}k^{(i)}(t)+\frac{N-1}{\sqrt{N}}\bar{L}^{(i)}(t)\right)\ket{\phi} \nonumber 
\\ &+& k^{(i)}(t)\ket{x_i}-\sum_{k=1}^{i-1}l_k^{(i)}\ket{x_k}-\sum_{k=i+1}^{N}\alpha^{(i)}(t)\ket{x_k}\Bigr\}
\nonumber \\ &=& \frac{2}{N} \bar{L}^{(i)}(t)(N-1)-\frac{2}{N}k^{(i)}(t)-\alpha^{(i)}(t)\nonumber 
\end{eqnarray}

Using Eqs.~(\ref{eq:k_L}) and (\ref{eq:alpha}) as well as the initial values of $\alpha(0)$, $k(0)$ and $\bar{L}(0)$ at the beginning of the Grover iterations as well as the number of iterations $t_f^{(i)}$, with a simple \textit{classical} iterative procedure we can compute the final values of $\alpha^{(i)}(t_f^{(i)})$, $k^{(i)}(t_f^{(i)})$ and $\bar{L}^{(i)}(t_f)$. 
This concludes one step of the embedding algorithm.

\subsection{Linking two consecutive steps of the distribution-encoding algorithm}

Let us see how to use the $\alpha^{i}(t_f^{(i)})$, $k^{i}(t_f^{(i)})$ and $\bar{L}^{i)}(t_f^{(i)})$ computed at the end of step $i$ to obtain the values $\alpha_{(i+1)}(0)$, $k^{(i+1)}(0)$ and $\bar{L}^{(i+1)}(0)$, which are needed in the following step $i+1$.

We first consider the global state at the end of the step $i$ before applying the $\vu*{F}_{i}$ operator to mark the state $\ket{x_i}$ whose amplitude has just been updated:
\begin{align*}
\ket{\psi(t_f^{(i)})}=\sum_{j=1}^{i-1}a_j\ket{x_j}\ket{0}_a+b_{i}\ket{\beta_{i}(t_f^{(i)})}\ket{1}_a
\end{align*}
where
\begin{align}
\label{eq:before}
\ket{\beta_{i}(t_f^{(i)})}=k^{(i)}(t_f^{(i)})\ket{x_{i}}+\sum_{k \neq {i}}l_k(t_f^{(i)})\ket{x_k}.
\end{align}
After applying $\vu*{F}_{i}$ we obtain the initial global state which will be the seed of the Grover iterations at the step $i+1$:
\begin{align}
\label{eq:after}
\ket{\psi(0)}=\sum_{k=1}^{i-1}a_k\ket{x_k}\ket{0}_a+a_{i}\ket{x_i}\ket{0}_a+b_{i+1}\ket{\beta_{i+1}(0)}\ket{1}_a,
\end{align}
where
\begin{align*}
\ket{\beta_{i+1}(0)}=k^{(i+1)}(0)\ket{x_{i+1}}+\sum_{k \neq i+1}l_k^{(i)}(0)\ket{x_k}
\end{align*}
In this new state, the coefficient $b_{i+1}$ ensures that $\ket{\beta_{i+1}(0)}$ is unit. By looking at the coefficient of $\ket{x_{i}}$ in both (\ref{eq:before}) and (\ref{eq:after}), it is easy to see that $a_{i}=b_{i}k^{(i)}(t_f^{(i)})$. In the same way, we can compare the coefficients that in (\ref{eq:after}) appear with the state $\ket{1}_a$ of the ancilla with the corresponding components in (\ref{eq:before}): 
\begin{eqnarray*}
b_{i+1}\ket{\beta_{i+1}(0)}\ket{1}_a & = & b_{i}\Bigl(\ket{\beta_{i}(t_f^{(i)})}-k^{(i)}(t_f^{(i)})\ket{x_{i}}\Bigr)\ket{1}_a= \\ &=& b_{i}\Bigl(\sum_{k \neq {i}}l_k^{(i)}(t_f^{(i)})\ket{x_k}\Bigr)\ket{1}_a
\end{eqnarray*}
It follows that:
\begin{align*}
\ket{\beta_{i+1}(0)}=\frac{b_i}{b_{i+1}}\sum_{k \neq i}l_k^{(i)}(t_f^{(i)})\ket{x_k}
\end{align*}
Finally we obtain that:
\begin{eqnarray*}
&& \alpha^{(i+1)}(0)=k^{(i+1)}(0)=\frac{b_i}{b_{i+1}}\alpha^{(i)}(t_f^{(i)}) \\
&& \bar{L}^{(i+1)}(0)=\frac{\sum_{k \neq i+1}l_k^{(i+1)}(0)}{N-1}=\frac{b_i}{b_{i+1}}\frac{\sum_{k \neq i+1}l_k^{(i)}(t_f^{(i)})}{N-1}= \\ && \quad = \frac{b_i}{b_{i+1}}\left(\bar{L}^{(i)}(t_f^{(i)})-\frac{\alpha(t_f^{(i)})}{N-1}\right)
\end{eqnarray*}

It is also possible to generalize these results in the case in which the states to be updated are superposition of more than one basis states. Let us assume that we have $r_i>1$. In this case the general relations for $\bar{K^{(i)}}(t)$ and $\bar{L^{(i)}}(t)$ read :
\begin{align*} 
\bar K^{(i)}(t)&=\bar{K}^{(i)}(0)\cos(wt)+\bar L^{(i)}(0)\sqrt{\frac{N-r_i}{r_i}}\sin(wt)\\
\bar L^{(i)}(t)&=\bar L^{(i)}(0)\cos(wt)-\bar{K}^{(i)}(0)\sqrt{\frac{r_i}{N-r_i}}\sin(wt).
\end{align*}
with $w=2\arcsin(\sqrt{r_i/N})$.
Observing that all the marked states have the same probability we conclude that the probability of a single state is $\bar K^{(i)}(t)^2$.
Moreover the expression for $\bar{L}^{(i+1)}(0)$ now is:
\begin{align*}
\bar{L}^{(i+1)}(0)=\frac{b_i}{b_{i+1}}\left(\frac{(N-r_i)\bar{L}^{(i)}(t_f^{(i)})-r_{i+1} \alpha^{(i)}(t_f^{(i)})}{N-r_{i+1}}\right),
\end{align*}
The other updating rules will be the same.
It is important to underline that in this general case it is necessary to previously know the number of states related to each oracle. In order to this a previous quantum counting procedure is needed.

\section{Complexity and Precision}
In order to compute the complexity of the algorithm, we start from the observation, derived in Ref.~\cite{GeneralizedGrover}, that the optimal number of Grover's iterations for a given step i is upper bounded by :
\begin{align*}
N_I^{(i)}=\frac{\frac{\pi}{2}-\arctan(\frac{\bar K(0)}{\bar L(0)}\sqrt{\frac{r_i}{N-r_i}})}{\arccos(1-2\frac{r_i}{N})}.
\end{align*}
Expanding $N_I^{(i)}$ to the leading-order in our working assumptions ($N>>1$), we obtain:
\begin{align*}
N_I^{(i)} \simeq -\frac{1}{2} \frac{\bar K(0)}{\bar L(0)}+\frac{\pi}{4}\sqrt{\frac{N}{r_i}}.
\end{align*}
From this expression, we can conclude that the initial conditions of the state can only reduce the optimal number of calls of Grover's operators because we are dealing only with positive amplitudes, and we have
$\bar K(0),\bar L(0) \ge 0$.
It is worth to note that in our case we want to set the amplitude of each action not to the maximum value but to the value that is determined by a probability distribution, therefore the number of Grover's iterations is typically much lower than the upper bound given above.
%Of course, this is an upper bound for the number of Grover's iterations referred to a generic action because for our goal reaching the number of optimal iterations for a step is a particular case.
%In particolare poichè il nostro obiettivo è approssimare una distribuzione per la maggior parte degli steps il numero di iterazioni di $\vu*{G}_i$ sarà di gran lunga inferiore a quella ottimale.
As explained in Sec.~\ref{secdue}, the number of times that the Grover procedure has to be executed is equal to the number of sub-intervals (J) chosen, and so the total complexity is: 
\begin{align*}
\textbf{O}\left(J \sqrt{N}\right)=\textbf{O}\left(\sqrt{N}\right),
\end{align*}
where we took into account that for a given state $s$, $N=|A_{s}|$, the complexity is equal to $\textbf{O}\left(\sqrt{|A_{s}|}\right)$.
Another important property to analyse is the precision of the algorithm. With precision we mean the sensibility in the increment of the probability of one action after one iteration (from t to t+1).
It can be quantified as follows:
%%\begin{widetext}
\begin{eqnarray}
\Delta P&=&|\bra{x_i}\bra{1}_a\ket{\psi(t+1)}|^2-|\bra{x_i}\bra{1}_a\ket{\psi(t)}|^2\\
&=&b_i^2\Bigl{(}\bar{K}(t+1)^2-\bar K(t)^2\Bigl{)}\nonumber\\
&=&b_i^2\Bigl{(}\bar K(0)^2[\cos(w(t+1))^2 -\cos(wt)^2]\nonumber
\\&+& \bar L(0)^2\bigl{(}\frac{N-r_i}{r_i}\bigl{)}[\sin(w(t+1))^2-\sin(wt)^2]\nonumber\\
&+&\bar K(0) \bar L(0) \sqrt{\frac{N-r_i}{r_i}} [\sin(2w(t+1))-\sin (2wt)]\Bigl{)}\nonumber.
\end{eqnarray}
%%\end{widetext}

Recalling that we assumed $|A_{s}|>>1$ and considering the upper bound case in which $t \sim \sqrt{|A_{s}|}$ and $r_1<<N$, we can expand $\Delta P$ to the leading-order in $1/|A_{s}|$, we have:
\begin{align*}
\Delta P \sim b_i^2\Bigl{(}&\bar K(0)^2\frac{1}{\sqrt{|A_{s}|}}+\bar L(0)^2|A_{ s}|\frac{1}{\sqrt{|A_{s}|}}+\\
& \bar K(0) \bar L(0) \sqrt{|A_{s}|}\Bigl{)}
\end{align*}

The final upper bound for the precision is obtained in the case $b_i=1$, $K(0) \sim |A_{s}|^{-1/2}$ and $\bar{L}(0) \sim |A_{s}|^{-1/2}$:
\begin{align*}
\Delta P \sim \frac{1}{\sqrt{|A_{s}|}}.
\end{align*}
\end{document}